\documentstyle[12pt]{article}
\def\k{\kappa}
\def\rr#1{(#1)}
\def\t{\hat t}
\def\tens{\otimes}
\def\bel{\begin{equation}\label}
\def\ee{\end{equation}}
\def\r#1{(\ref{#1})}
\def\poin{Poincar\'e }
\def\kdef{$\kappa$-deformed }
\def\k{\kappa}
\def\hx{\hat x}
\def\hp{\hat p}
\def\beq{\begin{eqnarray}}
\def\eeq{\end{eqnarray}}
\def\ba{\begin{array}}
\def\ea{\end{array}}
\def\lbl{\label}
\def\D{\Delta}
\def\<{\left<}
\def\>{\right>}
\newcounter{popnr}
\def\theequation{\thesection.\arabic{equation}}
\renewcommand{\theequation}{\arabic{section}.\arabic{equation}}
\newcommand{\alpheqn}{\setcounter{popnr}{\value{equation}}
		      \addtocounter{popnr}{1}
		      \setcounter{equation}{0}
\renewcommand{\theequation}{\arabic{section}.\arabic{popnr}\alph{equation}}}
\newcommand{\reseteqn}{\setcounter{equation}{\value{popnr}}
     \renewcommand{\theequation}
     {\arabic{section}.\arabic{equation}}}
\def\bl{\alpheqn}
\def\el{\reseteqn}

\normalbaselines
\catcode `\@=11
\@addtoreset{equation}{section}

\def\theequation{\arabic{section}.\arabic{equation}}
\def\section{\@startsection {section}{1}{\z@}{-3.5ex plus -1ex minus
     -.2ex}{2.3ex plus .2ex}{\normalsize\bf}}
\def\subsection{\@startsection{subsection}{2}{\z@}{-3.25ex plus -1ex minus
 -.2ex}{1.5ex plus .2ex}{\normalsize\bf}}
\def\@cite#1#2{${}^{\mbox{\scriptsize#1\if@tempswa , #2\fi}}$}
\def\thebibliography#1{\section*{References\markboth
  {REFERENCES}{REFERENCES}}\list
  {\arabic{enumi}.}{\settowidth\labelwidth{[#1]}\leftmargin\labelwidth
  \advance\leftmargin\labelsep
  \usecounter{enumi}}
  \def\newblock{\hskip .11em plus .33em minus -.07em}
  \sloppy
  \sfcode`\.=1000\relax}
 
\catcode `\@=12

\begin{document}

\begin{flushright}
OUTP-97-24P \\
hep-th/9706031
\end{flushright}

\vspace*{2.5cm}
\noindent
{ \bf $\kappa$-DEFORMED COVARIANT PHASE SPACE AND QUANTUM-GRAVITY
UNCERTAINTY RELATIONS}\vspace{1.3cm}\\\noindent
%\vspace*{2.5cm}
%\noindent
%{ \bf UNCERTAINTY RELATIONS FOR SPACE-TIME MEASUREMENTS AND
%$\kappa$-DEFORMED PHASE SPACE}\vspace{1.3cm}\\\noindent
\hspace*{1in}
\begin{minipage}{13cm}
Giovanni Amelino-Camelia $^{1}$, Jerzy Lukierski $^{2}$
and Anatol Nowicki$^{3}$
\vspace{0.3cm}\\
 $^{1}$ Theoretical Physics, University of Oxford,\\
\makebox[3mm]{ }1 Keble Road, Oxford OX1 3NP, UK\\
 $^{2}$ Dept. of Mathematical Sciences, Science Laboratories,\\
\makebox[3mm]{ } University of Durham, South Road, Durham DH1 3LE, UK\\
 $^{3}$ Institute of Physics, Pedagogical University,\\
\makebox[3mm]{ } pl. S\l{}owia\'nski 6, 65-029 Zielona G\'ora, Poland\\

\makebox[3mm]{ }
\end{minipage}

\vspace*{0.5cm}

\begin{abstract}
\noindent
We describe the deformed covariant phase space corresponding to
the so-called $\kappa$-deformation of $D=4$ relativistic symmetries,
with quantum ``time''
coordinate and Heisenberg algebra obtained according
to the Heisenberg double construction.
The associated modified uncertainty relations
are analyzed, and in particular it is shown that these relations are
consistent with independent estimates 
of quantum-gravity limitations on
the measurability of space-time distances.
\end{abstract}

% section 1
\section{\hspace{-4mm}.\hspace{2mm}Introduction}

\hspace*{1cm} Recently there has been a lot of interest (see,
{\it e.g.}, [1-10]) in the possibility that
classical ideas about space-time structure
might fail to describe physics
below some minimal length $l_{min}$.
In general relativity the metric is a dynamical
quantity, determined by the presence of matter and energy.
Inserting in the classical Einstein equations the
quantum-mechanical indeterminacy of energy the response
of space-time implies that below the Planck length
$l_{\min}\simeq 10^{-33} cm$ the classical concept of space-time
is not applicable.
In string theories, the appearance of minimal
distances follows from the analysis of string collisions at Planckian
energies, which are found to be characterized
by the following modified uncertainty relation (see [2])
\bel{1.1}
\Delta x \geq \frac{\hbar}{\Delta p} + \alpha G \Delta p\, ,
\ee
where $G=l_p^2$ is the gravitational coupling (Newton) constant
and $\alpha$ is a constant related with the string tension (Regge slope). 
The relation implies a lower bound on the
measurability of distances
\bel{1.2}
\min \Delta x \sim \sqrt{\hbar \alpha G } \, .
\ee

It is quite plausible that at very short distances some sort of
quantization of space-time coordinates
could provide
an algebraic abstraction of
this or similar measurability bounds.
Similar ideas have been discussed
since the late 1940s (see, {\it e.g.}, [11, 12])
but new possibilities appeared quite recently with  the
development of the theory of quantum groups and quantum deformations
of Lie algebras (see, {\it e.g.}, [13-17]).
The notion of quantum group as a Hopf algebra permits
to consider deformed symmetries;
in fact, the Hopf algebra  axioms
provide simultaneously an algebraic generalization of the definition
of Lie group as well as of Lie algebra.
The phase space containing the
coordinate and momentum sectors can be described in
the quantum-deformed case by the Heisenberg double construction
(see, {\it e.g.}, [18, 19]).  An important part of
the definition of the Heisenberg double is the duality
relation between quantum Lie groups and quantum Lie algebras, which
generalizes the Fourier transform relation between the ordinary
coordinate and momentum spaces.

The standard form of covariant fourdimensional Heisenberg commutation
relations, describing quantum-mechanical covariant phase space looks
as follows:

\bel{1.3}
[x_\mu,p_\nu] = i\hbar g_{\mu\nu}\,,	 \qquad g_{\mu\nu} =
\mathop{\rm diag} (-1,1,1,1).
\ee
Our aim here is to investigate the deformation of the phase space
commutation relations \r{1.3}
following from the so-called quantum $\k$-deformations [20-23]
of relativistic symmetries.
By using the algebraic scheme of Heisenberg doubles,
we introduce the $\k$-deformation
of covariant phase space relations~\footnote{The relations
characterizing this deformation were previously available only
in the conference reports [23, 24].},
and comment on some desirable features of such a deformation.
In particular, we investigate the implications of this
deformation for the measurability of space-time distances,
and, generalizing the results reported recently in [10],
we find overall consistency with earlier heuristic analyses
of measurability bounds in quantum gravity.
These findings provide support for the possibility that
the covariant $\k$-deformed phase space
give an effective description (at scales below the Planck scale)
of certain quantum-gravity or string-theory effects [10].

\section{Quantum Deformations of $D=4$ relativistic phase space}

The space-time coordinates $x_\mu$
($\mu=0, 1, 2, 3$)
can be identified with the translation sector of the Poincar\'e
group, and the fourmomenta $p_\mu$ ($\mu=0, 1, 2, 3$) are given by
the translation generators of the \poin algebra. 
In considering
quantum deformations of relativistic symmetries as describing
the modification of space-time 
structure one is lead to the study of the
possible quantum \poin groups.
The classification of quantum
deformations  of $D=4$ \poin groups in the
framework of Hopf algebras was given by
Podle\'s and Woronowicz ([26]; see also [27]) and provides
the most general
class of noncommutative space-time coordinates $\hx_\nu$
allowed by the quantum-group formalism. One obtains the following
algebraic relations (we put $h=c=1$)
\bel{2.1}
(R-1)_{\mu\nu}{}^{\rho\tau}(\hat x _\rho \hat x _\tau +\frac 1\kappa
T_{\mu\nu}{}^\rho \hat x_\rho +\frac1{\kappa^2}C_{\mu\nu})=0\,,
\ee
where
\begin{itemize}
\item the matrix $R$ satisfies the relation $R^2=1$ and
describes the allowed class of
quantum Lorentz groups;
\item $T_{\mu\nu}{}^\rho$, $C_{\mu\nu}$ are numerical dimensionless
tensors, satisfying suitable conditions [26].
\item $\kappa$ describes the fundamental mass parameter
\end{itemize} If we assume
that the quantum deformation does
not affect the nonrelativistic kinematics, i.e. we preserve the
nonrelativistic $O(3)$ rotations classical, the 
only consistent class of
noncommuting space-time coordinates is described by the relations
of the \kdef Minkowski
space (see [28, 23]), obtained from \r{2.1} by putting $R=\tau$
(classical Lorentz symmetry)  where $\tau$ is a flip operator
($\tau(a\tens b)=b\tens a$),
$T_{\mu\nu}^\rho
=\delta_\mu{}^0\delta_\nu{}^\rho-\delta_\nu{}^0\delta_\mu{}^\rho$
and $C_{\mu\nu}=0$, i.e. ($i=1,2,3$)
\bl
\beq
\lbl{2.2a}
[\hx_0,\hx_i]&=&\frac i \kappa \hx_i\,, \\{}
\lbl{2.2b}
[\hx_i,\hx_j]&=&0\,.
\eeq
\el
 The relations \r{2.2a}--\r{2.2b} describe the translation
sector of \kdef $D=4$ \poin group, and in particular \r{2.2b} 
reflects the classical nature of the space coordinates.

We choose the basis for $\kappa$-Poincar\'{e} algebra, with the
following $\kappa$-deformed Hopf algebra of fourmomentum
${\cal{P}}_{\kappa}$ $(k=1,2,3)$
\bl
\beq
[\hp_{0}, \hp_{k}] &=& 0 \lbl{2.4a} \\
\Delta(\hp_0) &=& \hp_0 \otimes 1 + 1\otimes \hp_0 \nonumber \\
\Delta(\hp_k) &=& \hp_k \otimes 1 + e^{{\hp_{0}}\over \kappa c}\otimes
\hp_k\lbl{2.4b}
\eeq
\el
and the antipode and counit is given by
\bel{2.4c}
S(\hp_\mu) = -\hp_\mu\qquad\qquad \epsilon(\hp_\mu)=0\,,
\ee
where the fundamental constant $c$ (light velocity) is inserted in
\r{2.4b}.

Using the duality relations with second fundamental constant $\hbar$
(Planck's constant)

\bel{2.5}
\<\hx_\mu, \hp_\nu\> = -i\hbar g_{\mu\nu}\qquad g_{\mu\nu} = (-1,1,1,1)
\ee
we obtain the noncommutative $\kappa$-deformed configuration space
${\cal{X}}_{\kappa}$ as a Hopf algebra with the following algebra and
coalgebra structure \bigskip
\bl
\beq
[\hx_{0}, \hx_{k}] &=& {{i\hbar}\over \kappa c}\hx_k\,,\qquad\qquad
[\hx_{k}, \hx_{l}] = 0\hfill\lbl{2.6a}\\[3mm]
\Delta(\hx_\mu ) &=& \hx_{\mu}\otimes 1 + 1\otimes \hx_{\mu}\,,
\lbl{2.6b}\\ S(\hx_{\mu}) &=& -\hx_{\mu}\qquad\qquad
\epsilon(\hx_{\mu}) = 0\hfill \lbl{2.6c}\eeq
\el
The $\kappa$-deformed phase space can be considered as the
Heisenberg double i.e a vector space  ${\cal{X}}_{\kappa}\otimes
{\cal{P}}_{\kappa}$ with product
\bel{2.7}
(x\otimes p)(\tilde{x}\otimes \tilde{p})=x(p_{(1)}\triangleright
\tilde{x})\otimes p_{(2)}\tilde{p}
\ee
where left action is given by
\bel{2.8}
p\triangleright x=\<p,x_{(2)}\>x_{(1)}
\ee
The product \r{2.7} can be rewritten as the commutators between
coordinates and momenta using the obvious isomorphism
$x\sim x\otimes 1$, $p\sim 1\otimes p$, which applied to
the case of $\kappa$-Poincar\'{e}
algebra provides the following relations
\bel{2.9}
\ba{rclrcl}
[\hx_k,\hp_l] &=&i \hbar \delta_{kl}\,,
\qquad&[\hx_k,\hp_0]&=&0\,,\\[3mm]
[\hx_0,\hp_k] &=& -{{i\hbar}\over \kappa c} \hp_k\,,
\qquad&[\hx_0,\hp_0]&=&-i\hbar\,. \ea
\ee
The set of relations \r{2.2a}--\r{2.4c} and \r{2.9} describes the
\kdef relativistic quantum phase space.

 From these relations follow the  modified covariant Heisenberg
uncertainty
relations. Introducing the dispersion of the observable
$a$ in quantum mechanical sense by
\bel{2.10a}
\Delta(a) \ = \ \sqrt{\<a^2\> - \<a\>^2}
\ee
we have
\bel{2.10b}
\Delta(a)\Delta(b)\geq {1\over 2}|\<c\>|\qquad\mbox{ where }\qquad
c=[a,b] \ee
We obtain $\kappa$-deformed uncertainty relations
\bl
\beq
\Delta(\t)\Delta(\hx_k)&\geq& {\hbar\over 2\kappa
c^2}|\<\hx_k\>|=\frac12\frac{l_\k}{c}\,\left|\<\hx_k\>\right|\,,\lbl{2.11a} 
\\\Delta(\hp_k) \Delta(\hx_l)&\geq& {1\over 2}\hbar\delta_{kl}
\lbl{2.11b}\,,\\ \Delta(\hat E) \Delta(\t)&\geq& {1\over 2}\hbar
\lbl{2.11c}\,, \\ \Delta(\hp_k) \Delta(\hat t)&\geq& {\hbar\over
2\kappa c^2}|\<\hp_k\>|=\frac12\frac{l_\k}{c}\, \left|\<p_k\>\right|
\lbl{2.11d}\,.\eeq \el
where $l_\k=\frac{\hbar}{\kappa c}$ describes the fundamental
length at which the time variable should already be considered
noncommutative. In the recent estimates $\kappa > 10^{12} GeV$
(see e.g.\ [30]) i.e.\ $l_\k
< 10^{-26}$cm; in particular one can put $\kappa$ equal to the
Planck mass what implies that $l_\k=l_p \simeq 10^{-33}$cm.

The relations \rr{2.3a}-\rr{2.3b} describe the fourmomentum sector
of the \kdef \poin algebra, written in the bicrossproduct
basis [22, 23] with classical Lorentz algebra.
\bl
\bel{2.13a}
[M_{\mu\nu}, M_{\rho\tau}] = i(g_{\mu\rho}M_{\nu\tau}+
g_{\nu\tau}M_{\mu\rho}-g_{\mu\tau}M_{\nu\rho}
-g_{\nu\rho}M_{\mu\tau})
\ee
and the following deformed covariance relations ($M_i=
\frac12\epsilon_{ijk} M_{jk}, N_i=M_{i0}$)
\bel{2.13b}
\ba{rcl}
[M_i, P_j]&=&i\epsilon_{ijk}P_k\,\qquad [M_i,P_0]\,=\,0\\{}
[N_i,P_j]&=& -i\delta_{ij}\left[\frac{\kappa c}2
(1 -e^{2\frac{P_0}{\k c}} ) +
\frac1{2\k c} (\vec P)^2\right] + \frac {i}{\k c} P_i P_j\,,\\{}
[N_i,P_0]&=&iP_i\,,
\ea
\ee
\el
where the translation generators $P_\mu$ should be identified with the 
$\kappa$-relativistic phase space coordinates $\hat p_\mu$. 

The \kdef mass Casimir  takes the form
\bel{2.14}
C_2^\k = {1\over c^2}\vec P^2 e^{-\frac{P_0}{\k c}} - (2\k \sinh
\frac{P_0}{2\k c} )^2 = -M^2\,,
\ee
where $M$ denotes the $\k$-invariant mass parameter. In particular
for $M=0$ (\kdef photons) from \r{2.14} one obtains that
\bel{2.15}
P_0=\k c \ln (1+\frac{|\vec P|}{\k c }) = |\vec P|
-\frac{|\vec P|^2}{2 \k c} + O(\frac1{\k^2})
\ee
and in particular the velocity formula for massless \kdef quanta
looks as follows ($E=cP_0$)
\bl
\bel{2.16a}
V_i = c\frac{\partial P_0}{\partial P_i} = \frac{c}{1+
\frac{|\vec P|}{\k c}} \frac{P_i}{|\vec P|} 
\ee
or
\bel{2.16b}
V=|\vec{V}|=\frac {c}{1+\frac{|\vec P|}{\k c}} = c-
\frac{|\vec P|}{\k} +O(\frac 1 {\k^2})
\ee
\el

%\begin{comment}
%{\bf DISCUSSION OF DEFORMED KLEIN-GORDON EQUATION (OF
%GREAT INTEREST%FOR THE PHYSICS ORIENTED AUDIENCE) COULD BE INSERTED HERE}
%\end{comment}

This three-momentum-dependent (i.e. energy-dependent)
``speed of light'' is a completely novel phenomenon that arises
in the formalism here considered. Interestingly, it has the same
functional form (upon appropriate identification between
$\kappa$ and the string scale) as the energy-dependent speed of light
recently discussed [9]
in the non-critical  (``Liouville'') string  literature.
Both in the $\kappa$-Poincar\'{e} and in the string theory
contexts the deviation from ordinary physics,
while very significant at the conceptual level,
is rather marginal from the phenomenological viewpoint.
For example, for photons of energies of order 1 GeV
the Eq. \r{2.16b} entails a
minuscule $10^{-19} c$ correction with respect to the ordinary
scenario with constant speed of light.
As discussed in greater detail in [9], at least when
$\kappa$ is identified with the Planck scale, the Eq.\r{2.16b}
is completely consistent with available experimental data.
As manifest in the 
relations \r{2.11a}-\r{2.11d}, the $\k$-modifications of the covariant 
Heisenberg commutations relations are of quantum mechanical nature, i.e. 
proportional to the Planck constant $\hbar$. This suggests that the 
$\k$-deformation (together with its exotic energy-dependent
speed of light)
can be related with the quantum corrections to the 
classical dynamics of space-time.

\section{Measurement process and covariant \kdef phase space}

Recently one of the present authors has presented
(see [8]) heuristic quantum-gravity arguments
(based on combining quantum mechanics with general relativity)
indicating that the measurability of distances
may be bound by a quantity that grows with the time required
by the measurement procedure, as needed for the decoherence
mechanism discussed in [4].
Specifically, by observing
that gravitational effects prevent one from relying
on the availability of {\it classical}
agents for the measurement procedure
(since the limit of infinite mass
leads to inconsistencies associated with the formation
of horizons), the following bound
is found for the measurability of a distance $L$ [8]:
\bel{qgboundgac}
\min \left[ \D L \right] \sim l_p \sqrt{ c T \over s} \sim
l_p\sqrt{ L \over s }
% \sim \sqrt{ {L L_P^2 \over s}}
~,
\ee
where $l_p$ is the Planck length,
$s$ is a length scale characterizing
the spatial extension of the devices ({\it e.g.}, clocks)
used in the measurement, $T$ is the
time needed to complete the procedure of measuring $L$,
and on the right-hand-side we used the fact that typically $T \sim L$.
Further analyses of this type of measurability bound have relied
again on the gravitational effects associated to
macroscopic devices or on the study of the
light probes exchanged during the measurement; in particular,
the dynamics of the light probes was shown to lead to a
measurability bound of type \r{qgboundgac}
in the framework of Liouville noncritical string theories with the
target time identified with the Liouville mode [9].

In this section we analyze the measurement of the distance $L$
between two bodies as it results from a plausible physical
interpretation of the uncertainty relations \r{2.11a}-\r{2.11d}.
Like the related studies [4, 8, 9]
we consider the procedure of measurement of distances
set out by Wigner [29, 30],
which relies on the exchange of a light
probe/signal between the bodies.
The distance is therefore measured as $L = c \, T/2$,
where $T$ is the time
spent by the probe to go from one body to the other and return.
In general the quantum mechanical nature of the probe
introduces uncertainties in the measurement of $L$,
and in particular one finds that~\footnote{Of course
there are other contributions to $\Delta L$ ({\it e.g.}, coming
from the quantum mechanical nature of the other devices used
in the experiment [8]); however, since they obviously
contribute additively to the total uncertainty
in the measurement of $L$,
these uncertainties could only make stronger
the bound derived in the following.}
\bel{3.1}
\D L \geq \Delta x + c \, \Delta t + T \Delta v
\ee
where $\Delta x$ and $\Delta t$ are the uncertainties
on the space-time position~\footnote{As implicit in the
terminology here adopted, the Wigner measurement procedure is
essentially one-dimensional, and the only relevant spatial
coordinate is the one along the axis passing through
the bodies whose distance is being measured.}
of the probe at time $T$,
while $\Delta v$ is the uncertainty on the velocity of the probe.

The analysis of these uncertainties within ordinary quantum mechanics
would be 
 trivial since 
$\Delta x$, $\Delta t$ and  $\Delta v$ are
not correlated in that framework.
The $\kappa$ deformation induces such correlations.
In particular,
concerning the correlation between 
$\Delta x$ and
$\Delta t$ we observe that \r{2.11a} implies 
(interpreting the $x$ on the right-hand-side of \r{1.3}
as the distance travelled by the probe)
\bel{3starb}
\Delta t \geq \frac{\hbar L}{2 \kappa c^2\, \Delta x} 
\, .
\ee
Moreover,  if the probe is massless with modified velocity \r{2.16b} one
finds that\bel{deltavofE}
|\Delta v| \sim  \frac{\Delta P}{\k}  \sim \frac{\hbar}{2 \k \, \D x}\,,
\ee
where on the right-hand-side we used \r{2.11b}.

Using \r{3starb} and \r{deltavofE}
one can rewrite \r{3.1} as
\bel{3star}
\D L \geq \Delta x +
\frac{\hbar L}{2 \kappa c\, \Delta x} +
\frac{\hbar T}{2 \kappa \, \Delta x}  
~.
\ee

Taking into account that $L = c \, T/2$
 one finds that the minimal value of $\D L$ is obtained if
$(\D x)^2= \frac54 \frac{\hbar L}{\k c}$ and this implies that the
minimal uncertainty in the measurement of the distance $L$ is
\bel{3.6}
\min[\D L] \sim \sqrt{\frac{\hbar L}{\kappa c}}
\ee
Remarkably this result reproduces (up to appropriate
mapping between the scale $\kappa$ and the scales $l_p$ and $s$)
the relation \r{qgboundgac} that was derived within a completely
independent analysis of quantum-gravity effects.
The preliminary investigation reported
in [10], which
considered only the \kdef Minkowski coordinates sector,
had raised the possibility that this might be the case;
however, it is quite non-trivial
that the present analysis taking into account the structure
of the full covariantly \kdef phase space
ultimately leads to \r{3.6}.
The form of the $\kappa$-deformation advocated here
plays a rather central role in obtaining this result;
in fact, the relation \rr{2.16} ensures that the third term on the
right-hand side of Eq.~\r{3.1} (which was not considered in [10])
is of the same order (once the uncertainty relations are taken
into account) as the first term.

The deformation of phase space here considered has
also implications for the analysis of macroscopic devices
in the measurement, but this does not lead to additional bounds to
the measurability of $L$. To illustrate this,
let us consider for example the clock used in the measurement.
The uncertainties in the time indicated by the clock,
the position of the clock,
and the velocity of the clock affect the measurement of $L$
according to
\bel{3.1bis}
[\D L]_{clock} \sim \Delta x_{clock} + v_{clock} \, \Delta t_{clock}
+ T \, \Delta v_{clock}
~.
\ee
Again the $\kappa$ deformation induces correlations
between $\Delta x_{clock}$, $\Delta t_{clock}$, $\Delta v_{clock}$;
in fact, using
the \kdef uncertainty relations \r{2.11a}-\r{2.11d}
(and considering an ideal non-relativistic clock with $p = M v$)
one finds that
\bel{3.12}
\D v_{clock} 
\geq \frac{\hbar |v_{clock}|}{2\k c^2 \D t_{clock}} 
\ee
and 
\bel{3starbclock}
\Delta x_{clock} \geq 
\frac{\hbar v_{clock} T}{2 \kappa c^2\, \Delta t_{clock}} 
\, .
\ee

Therefore the contribution from the quantum-mechanical uncertainty
in the kinematics of the clock satisfies
\bel{3starbis}
[\D L]_{clock} 
\geq \frac{\hbar v_{clock} \, T}{2 \kappa c^2 \, \Delta t_{clock}} 
+ v_{clock} \Delta t_{clock}
+ \frac{\hbar v_{clock} \, T}{2\kappa c^2 \, \Delta t_{clock}}
\ee
For a given $v_{clock}$ one finds 
\bel{3.5bis} 
\min [\D L]_{clock} \sim v_{clock} \sqrt{\frac{\hbar T}{\kappa c^2}}
=v_{clock}\sqrt{\frac {l_\k}{c} T}\,,
\ee
whereas, of course, in ordinary quantum mechanics 
is $\min [\D L]_{clock} = 0$.
However, Eq.~\r{3.5bis}
does not lead to a genuine measurability bound,
since the observer would naturally prepare the system
with $v_{clock} = 0$.

It appears therefore that the covariant $\kappa$-deformation of phase
space leads to a measurability bound in agreement with some of
the heuristic quantum-gravity measurement analyses, and that this
bound emerges from the nature of the kinematics of the light probe,
whereas the kinematics of macroscopic devices does not
contribute to the bound.

\section{Closing Remarks}

The covariant \kdef relativistic symmetries
here considered, and the associated
covariant $\kappa$-deformation of the
Heisenberg algebra \r{2.9},
has several appealing properties as a candidate for the 
high-energy modification of classical relativistic symmetries.
As a dimensionful deformation it is relevant only
to the description of processes characterized by energies
of order $\kappa$ or higher.
In addition, in an appropriate sense, it provides 
a rather moderate (at least in comparison
with some of its alternatives) deformation of classical 
relativistic symmetries, which in particular reflects the 
reasonable expectation that, if any of the space-time 
coordinates is to be special, the special coordinate should be time.
(Interestingly this intuition appears to be also realized
in certain approaches to string theory, see {\it e.g.} [32].)

In conclusion, the fact that we have provided some evidence that the bounds
on the measurability of distances associated with the
uncertainty relations characterizing the \kdef covariant Heisenberg
algebra \r{2.2a}-\r{2.2b} and \rr{2.9} are consistent with
independent analyses of such measurability bounds in quantum
gravity suggests that the $\kappa$-deformation might
provide an effective description at ultra-short distances (perhaps
just above the
Planck length) of certain quantum-gravity effects.

\end{document}